\documentclass[twocolumn,showpacs,preprintnumbers,amsmath,amssymb]{revtex4}

\usepackage{graphicx}
\usepackage{dcolumn}
\usepackage{bm}
\usepackage{color}

\begin{document}


\title{Effective and Efficient Similarity Index for Link Prediction of Complex Networks}

\author{Linyuan L\"{u}$^{1}$}
\author{Ci-Hang Jin$^{1}$}
\author{Tao Zhou$^{1,2}$}
\email{zhutou@ustc.edu} \affiliation{$^1$Department of Physics,
University of Fribourg, Chemin du Muse, Fribourg 1700, Switzerland
\\ $^2$Department of Modern Physics, University of Science and
Technology of China, Hefei 230026, P.R. China}

\begin{abstract}
Predictions of missing links of incomplete networks like
protein-protein interaction networks or very likely but not yet
existent links in evolutionary networks like friendship networks in
web society can be considered as a guideline for further experiments
or valuable information for web users. In this paper, we introduce a
local path index to estimate the likelihood of the existence of a
link between two nodes. We propose a network model with controllable
density and noise strength in generating links, as well as collect
data of six real networks. Extensive numerical simulations on both
modeled networks and real networks demonstrated the high
effectiveness and efficiency of the local path index compared with
two well-known and widely used indices, the \emph{common neighbors}
and the \emph{Katz index}. Indeed, the local path index provides
competitively accurate predictions as the Katz index while requires
much less CPU time and memory space, which is therefore a strong
candidate for potential practical applications in data mining of
huge-size networks.

\end{abstract}


\maketitle

\section{Introduction}
Many complex systems can be well described by networks where nodes
present individuals or agents, and links denote the relations or
interactions between nodes. Complex network is therefore becoming an
useful tool in analyzing a wide range of complex systems. Recently,
the understanding of structure, evolution and function of networks
has attracted much attention from physics community
\cite{Albert2002,Dorogovtsev2002,Newman2003,Boccaletti2006,Costa2007}.
Another important scientific issue relevant to network analysis,
namely the \emph{Information Retrieval}
\cite{Salton1983,Salton1989}, however, received less attention.
Originally, Information Retrieval aims at finding material of an
unstructured nature that satisfies an information need from large
collections \cite{Manning2008}. It can be also viewed as dealing
with prediction of links between words and documents, and is now
further extended to standing for a number of problems on link mining
\cite{Getoor2005}. Actually, link prediction problem is a
long-standing challenge in modern information science, and a lot of
algorithms have been proposed based on Markov chains and machine
learning processes by computer science community
\cite{Sarukkai2000,Popescul2003,JZhu2002,Bilgic2007,Yu2007}.
However, their works have not caught up the current progress of the
study of complex networks, especially they lack serious
consideration of the structural characteristics of networks which
may indeed provide useful information and insights for link
prediction.

The problem of link prediction aims at estimating the likelihood of
the existence of a link between two nodes, based on observed links
and the attributes of nodes. It can be categorized into two classes:
One is the prediction of missing links in sampling networks, such as
the food webs and the world wide webs; the other is the prediction
of links that may exist in the future of evolving networks, like the
on-line social networks. In addition, the link prediction algorithms
(or other algorithms based on similar techniques) can also be
applied to solve the link classification problem in partially
labeled networks \cite{Holme2005,Gallagher2008}, such as the
prediction of protein functions \cite{Holme2005} and to distinguish
the research areas of scientific publications \cite{Gallagher2008}.

Up to now, most of the algorithms are designed according to the
definition of node similarity. Node similarity can be defined just
using the essential attributes of nodes, namely two nodes are
considered to be similar if they have many common features
\cite{Lin1998}. Another group of similarity indices is based solely
on the network structure, which is called structural similarity and
can be further classified as node-dependent, path-dependent and
mixed methods. An introduction and comparison of some similarity
indices is presented in Ref. \cite{Kleinberg2007} in which the
\emph{Common Neighbors} \cite{Lorrain1971}, \emph{Jaccard
coefficient} \cite{Jaccard1901}, \emph{Adamic-Adar Index}
\cite{Adamic2003} and \emph{Preferential Attachment}
\cite{Barabasi1999} are classified to be the node-dependent indices,
while \emph{Katz Index} \cite{katz1953}, \emph{Hitting Time}
\cite{Gobel1974}, \emph{Commute Time} \cite{Fouss2007}, \emph{Rooted
PageRank} \cite{Brin1998}, \emph{SimRank} \cite{Widom2002} and
\emph{Blondel Index} \cite{Blondel2004} are classified to be the
path-dependent indices. Besides, Leicht, Holme and Newman proposed a
measure to quantify the node similarity based on the assumption that
two nodes are similar if their immediate neighbors in the network
are themselves similar \cite{Leicht2006}. This leads to a
self-consistent matrix formulation of similarity that can be
evaluated iteratively using the adjacency matrix. This similarity
index can also be considered as a candidate for accurate link
prediction.

Besides the similarity-based prediction algorithms, some more
complicated methods are proposed recently. Clauset, Moore and Newman
proposed an algorithm based on the hierarchical network structure
\cite{Newman2008Nature,Redner2008}. Firstly, they use a hierarchical
random graph to statistically fit the real network data. Then the
dependence of the lateral-connection probability on the depth of the
nodes in the hierarchy can be inferred. Finally, one can predict the
missing links of the network according to the lateral-connecting
probability by ranking them in the descending order. Furthermore,
many efforts have been done for designing the recommender systems
\cite{Adomavicius05}. Actually, the process of recommending items to
a user can be considered as the prediction of missing links in the
user-item bipartite network \cite{Zhou2009}. Especially, physicists
have recently proposed some information recommendation algorithms
based on physical processes, such as energy diffusion
\cite{Zhou2007,Zhou2008,Zhang2007a} and heat conduction
\cite{Zhang2007b}. Although the relevant issue has not been fully
explored, it highlights a possibility to improve the accuracy and
efficiency of link prediction algorithms by applying classical
physics dynamics.

There are many difficulties for the studies of link prediction. One
is the sparsity of the target networks
\cite{Getoor2003,Madadhain2005,Rattigan2005}, which leads to a
serious problem that the prior probability of a link is typically
quite small, resulting in large difficulties in building statistical
models. The other problem is the huge size of real systems that
requires highly efficient algorithms. However, the complexity of
computational time and memory, being a crucial factor in real
applications, has not been systematically investigated. Generally
speaking, the accuracy of an algorithm and its computational
complexity have positive correlation, namely higher accuracy usually
implies higher complexity. Note that, any highly accurate algorithm
will become meaningless if the consuming time or memory is
unacceptable. Therefore, designing an accurate and fast algorithm is
a big challenge, especially for sparse and huge networks.

In this paper, we introduce a so-called \emph{local path index} to
characterize the node similarity. Extensive numerical simulations on
both modeled networks and real networks demonstrate that this
similarity index is simultaneously highly effective (its prediction
accuracy is much higher than the \emph{common neighbors}, and
competitive with the Katz index) and highly efficient (the time and
space required to compute it are much less than those for the Katz
index). Especially, when the network is huge, the local path index
shows great advantage compared with the Katz index since computing
the latter asks for a CPU time scaling as cube of the network size
while computing the former requires a linear CPU time as the network
size. We therefore think this local path index is a strong candidate
for potential practical applications in data mining of huge-size
complex networks.

\section{Method}
Considering an unweighted undirected simple network $G(V,E)$, where
$V$ is the set of nodes and $E$ is the set of links. The multiple
links and self-connections are not allowed. For each pair of nodes,
$x,y\in V$, we assign a score, $s_{xy}$. Since $G$ is undirected,
the score is supposed to be symmetry, say $s_{xy}=s_{yx}$. All the
nonexistent links are sorted in decreasing order according to their
scores, and the links in the top are most likely to exist. In this
paper, we adopt the simplest framework, that is, to directly set the
similarity as the score, so the higher score means the higher
similarity, and vice versa. In some link prediction algorithms, the
scores may be not directly related to a certain similarity
measurement, but describe the existence likelihood of links
\cite{Sarukkai2000,Popescul2003,JZhu2002,Bilgic2007,Yu2007,Newman2008Nature},
and in some other algorithms, scores may be generated by an
integration of some similarities of node pairs in the neighborhood
of the target links, such as the collaborative filtering method
\cite{Huang2005}.

\begin{figure}
\begin{center}
\center \includegraphics[width=7cm]{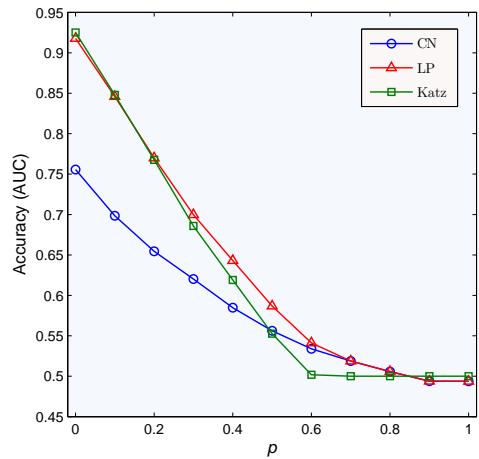} \caption{(Color
online) Prediction accuracy vs. the strength of randomness for three
similarity indices: CN (circles), LP (triangles) and Katz index
(squares). The network size, $N=1000$, and the degree, $k=10$, are
fixed. Each data point is obtained by averaging over 10 independent
realizations. When approaching the purely random case, $p=1$, the
accuracies of CN and LP go below 0.5, which is an artifact of the
specific constrain on identical degree. That is, in the purely
random case, two unconnected nodes with higher degrees in the
training set are of less probability to be connected in the probe
set since the total degree is identical for every node, however,
they generally have more common neighbors and thus higher
similarity.}\label{Noise_AUC}
\end{center}
\end{figure}
\begin{figure}
\begin{center}
\center \includegraphics[width=7cm]{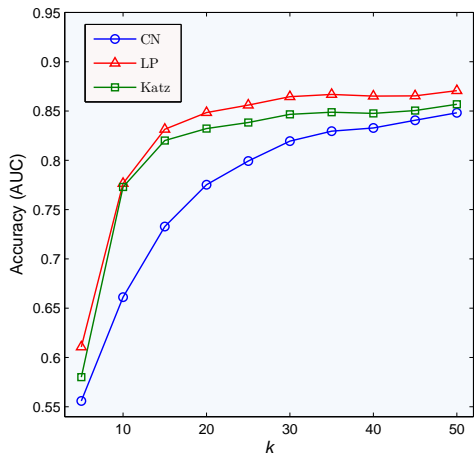} \caption{(Color
online) Prediction accuracy vs. network density for three similarity
indices: CN (circles), LP (triangles) and Katz index (squares).
Since in this model, every node has the same degree, we therefore
directly use degree to denote the network density. The network size,
$N=1000$, and the strength of randomness, $p=0.2$, are fixed. Each
data point is obtained by averaging over 10 independent
realizations.} \label{Degree_AUC}
\end{center}
\end{figure}

In this paper, we compare the prediction accuracies and
computational complexity of three similarity indices: \emph{Common
Neighbors} (CN), \emph{Katz Index} and a newly proposed similarity
index, namely \emph{Local Path Index} (LP index or LP for short).
Their definitions and relevant motivations are introduced as
follows:

(i)\emph{Common Neighbors}, which is also called \emph{structural
equivalence} in Ref. \cite{Lorrain1971}. In common sense, two nodes,
$x$ and $y$, are more likely to form a link in the future if they
have many common neighbors. For a node $x$, let $\Gamma(x)$ denote
the set of neighbors of $x$. The simplest measure of the
neighborhood overlap is the directed count:
\begin{equation}
s_{xy}=|\Gamma(x)\cap \Gamma(y)|,
\end{equation}
where $|Q|$ is the cardinality of the set $Q$. It is obvious that
$s_{xy}=(A^2)_{xy}$, where $A$ is the adjacency matrix, in which
$A_{xy}=1$ if $x$ and $y$ are directly connected and $A_{xy}=0$
otherwise. Note that, $(A^2)_{xy}$ is also the number of different
paths with length 2 connecting $x$ and $y$. Newman
\cite{Newman2001a} used this quantity in the study of collaboration
networks, showing the correlation between the number of common
neighbors and the probability that two scientists will collaborate
in the future. Some more complicated measures, such as \emph{Salton
Index} \cite{Salton1983}, \emph{Jaccard Index} \cite{Jaccard1901},
\emph{S{\o}rensen Index} \cite{Sorensen1948} and \emph{Adamic-Adar
Index} \cite{Adamic2003}, can also be categorized into CN-based
measures. However, recently, extensive empirical analysis has
demonstrated that the simplest CN (i.e., Eq. (1)) performs even
better than those complicated variants
\cite{Kleinberg2007,Linyuan2009}. Therefore, we here select CN as
the representative of all CN-based measures. Although CN consumes
little time and performs relatively good among many local indices,
due to the insufficient information, its accuracy can't catch up
with the measures based on global information. One typical example
is the \emph{Katz Index} \cite{katz1953}.

(ii) \emph{Katz Index}. This measure is based on the ensemble of all
paths, which directly sums over the collection of paths and
exponentially damped by length to give the short paths more weights.
The mathematical expression reads
\begin{equation}
s_{xy}=\sum^{\infty}_{l=1}\beta^l\cdot|paths^{<l>}_{xy}|,
\end{equation}
where $paths^{<l>}_{xy}$ is the set of all paths with length $l$
connecting $x$ and $y$, and $\beta$ is a free parameter controlling
the weights of the paths. Obviously, a very small $\beta$ yields a
measure close to CN, because the long paths contribute very little.
The $S$ matrix can be written as $(I-\beta A)^{-1}-I$. Note that,
$\beta$ must be lower than the reciprocal of the maximum of the
eigenvalues of matrix $A$ to ensure the convergence of Eq. (2).

(iii)\emph{Local Path Index}. To provide a good tradeoff of accuracy
and complexity, we here introduce an index that takes consideration
of local paths, with wider horizon than CN. It is defined as
\begin{equation}
S=A^2+\epsilon A^3,
\end{equation}
where $S$ denotes the similarity matrix and $\epsilon$ is a free
parameter. Clearly, this measure degenerates to CN when
$\epsilon=0$. And if $x$ and $y$ are not directly connected (this is
the case we are interested in), $(A^3)_{xy}$ is equal to the number
of different paths with length 3 connecting $x$ and $y$. Although it
needs more information than CN, it is still a local measure of
relatively lower complexity than global ones.

Choosing these three indices for comparison is because they all can
be classified to path-dependent similarities with unified form as
$s_{xy}=\sum\beta^l\cdot|paths^{<l>}_{xy}|$, where for CN, $l=2$;
for LP, $l=2,3$; and for Katz, $l=1,2,3,\cdots,\infty$. Since we are
only interested in the indirectly connected node pairs, Katz Index
can be treated as a measure considering $l=2,3,\cdots,\infty$. Note
that, all these three indices are used to quantify the structural
equivalence, with an latent assumption that the link itself
indicated a similarity between two endpoints (see, for example, the
Leicht-Holme-Newman index \cite{Leicht2006} and transferring
similarity \cite{Sun2009}). An issue worth future exploration is
whether a certain similarity measure on \emph{regular equivalence}
(see Ref. \cite{White1983} for the mathematical definition of
regular equivalence and Ref. \cite{Holme2005} for a recent
application on the prediction of protein functions) can provide
better predictions.

\begin{figure}
\begin{center}
\center \includegraphics[width=9cm]{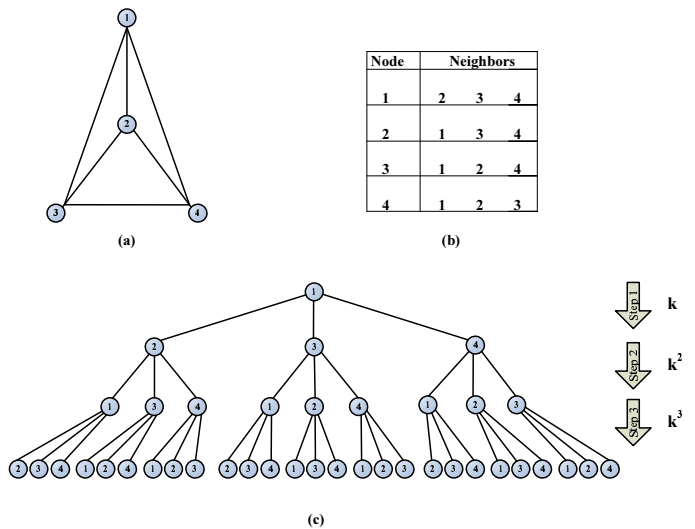} \caption{(Color
online) An illustration of time complexity in calculating CN and LP
indices. (a) A fully connected network with four nodes as the
example. (b) Lists of the neighborhood of each node. (c) Process of
how to determine all the similarities relevant to node
1.}\label{Time}
\end{center}
\end{figure}

To test the algorithmic accuracy, the observed links, $E$, is
randomly divided into two parts: the training set, $E^T$, is treated
as known information, while the probe set, $E^P$, is used for
testing and no information in the probe set is allowed to be used
for prediction. Clearly, $E=E^T\cup E^P$ and $E^T\cap
E^P=\varnothing$. In this paper, the training set always contains
90\% of links, and the remaining 10\% of links constitute the probe
set. We use a standard metric, \emph{area under the receiver
operating characteristic} (ROC) curve \cite{Hanely1982}, to quantify
the accuracy of prediction algorithms. In the present case, this
metric can be interpreted as the probability that a randomly chosen
missing link (a link in $E^P$) is given a higher score than a
randomly chosen nonexistent link (a link in $U\setminus E$, where
$U$ denotes the universal set). In the implementation, among $n$
times of independent comparisons, if there are $n'$ times the
missing link having higher score and $n''$ times the missing link
and nonexistent link having the same score, we define the
\emph{accuracy} as $\frac{n'+0.5n''}{n}$. If all the scores are
generated from an independent and identical distribution, the
accuracy should be about 0.5. Therefore, the degree to which the
accuracy exceeds 0.5 indicates how much better the algorithm
performs than pure chance. Readers are encouraged to see the Refs.
\cite{Geisser1993,Herlocker2004} for more information about how to
evaluate the accuracy of prediction algorithms.

\section{Model}
In this section, we compare the three similarity indices in modeled
networks with controllable density and randomness. Although the real
networks have complex structural properties \cite{Costa2007}, such
as the community structure, the mixing pattern and the rich-club
phenomenon, as a start point, we only consider a very simple model,
and to eliminate the effect of degree heterogeneity, we assume that
every node has an identical degree, $k$. In this model, each node is
characterized by a 10-dimensional vector with each element a
randomly selected real number in the interval $(-1,1)$. This vector
represents the node's intrinsic features, such as the attributes of
an object and the profiles of a person. Two nodes are considered to
be similar and thus of high probability to connect to each other if
they share many close attributes. Therefore, we define the
\emph{intrinsic similarity} between two nodes as the scalar product
of the corresponding vectors, namely
\begin{equation}
s^I_{xy}=\vec{f_{x}}\cdot{\vec{f_{y}}}=s^I_{yx},
\end{equation}
where $\vec{f_{x}}$ is the vector of node $x$, and the superscript
emphasizes that this similarity is intrinsic and can not be observed
in the real systems.

Given the network size, $N$, and the degree of each node, $k$, this
model starts with an empty network but $N$ nodes, that is, each node
is of degree zero. At each time step, a node with the smallest
degree is randomly selected (generally, there are more than one node
having the smallest degree). Among all other nodes whose degrees are
smaller than $k$, this selected node will connect to the most
similar node with probability $1-p$, while a randomly chosen one
with probability $p$. This process will terminate when all nodes are
of degree $k$. The parameter $p\in [0,1]$ represents the strength of
randomness in generating links, which can be understood as noise or
irrationality that exists in almost every real system.

\begin{figure}
\begin{center}
\center \includegraphics[width=7cm]{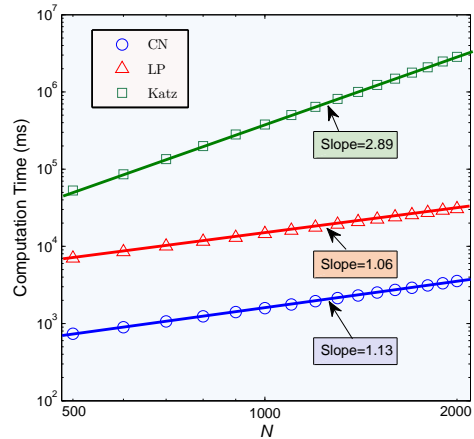} \caption{(Color
online) A log-log plot about how the computational time (in
microsecond) depends on the network size for three indices, CN
(circles), LP (triangles) and Katz (squares). The node degree,
$k=10$, and the strength of randomness, $p=0.2$, are fixed. Each
data point is obtained by averaging over 10 independent
realizations. All computations were carried out in a desktop
computer with a single Intel (R) Xeon (TM) processor (3.00 GHz) and
2GB EMS memory.}\label{N_size}
 \end{center}
\end{figure}

\begin{figure}
\begin{center}
\center \includegraphics[width=7cm]{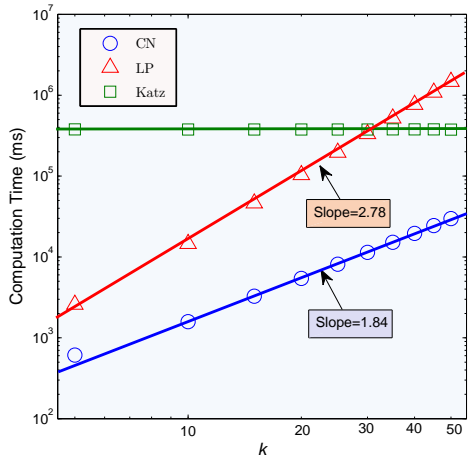} \caption{(Color
online) A log-log plot about how the computational time (in
microsecond) depends on the node degree for three indices, CN
(circles), LP (triangles) and Katz (squares). The network size,
$N=1000$, and the strength of randomness, $p=0.2$, are fixed. Each
data point is obtained by averaging over 10 independent
realizations. The hardware environment is the same as what we stated
in the caption of Figure 4.}\label{Degree}
 \end{center}
\end{figure}

In Fig. 1 and Fig. 2, we report the comparison of algorithmic
accuracy for those three similarity indices. Data points are
corresponding to the optimal values of $\beta$ (for Katz index) or
$\epsilon$ (for LP index) subject to the highest accuracies.
Clearly, both Katz index and LP index perform remarkable better than
the simple CN index. As shown in Fig. 1, when the strength of
randomness/noise is weak, LP index gives competitive result as Katz
index, while for highly noisy cases, LP index performs even better.
Whatever the similarity index, a link prediction algorithm is
expected to give higher accuracy for a denser network, which is in
accordance with what observed in Fig. 2. In the area with lacking
information (i.e., small $k$) or rich information (i.e., large $k$),
LP index performs slightly better than CN index, while in the middle
with typical degree as the real networks, LP index can perform much
better than the CN index.

The reason why the CN index performs remarkably poorer than LP index
is that the probability that two node pairs are assigned the same
similarity by CN is high. That is to say, CN index is less
distinguishable, especially in the relatively sparse networks. For
example, in the case $N=1000$, $p=0$ and $k=10$, there are about
$5\times 10^5$ node pairs, 94.01\% of which are assigned zero score,
and for all non-zero scores, 79.87\% are $1$. As shown later, the
real cases may be even worse, for instance, in a router-level
Internet with 5022 nodes, 99.59\% of node pairs are assigned zero
score by CN, while for all those non-zero scores, 91.11\% of which
are assigned score one. The additional information involving the
next nearest neighbors introduced by LP index can make the
similarities much more distinguishable, thus remarkably enhance the
accuracy. Note that, if the maximal number of paths with length
three connecting two arbitrary nodes is $P_{max}$, any $\epsilon$ in
the interval $(0,\frac{1}{P_{max}})$ will give out exactly the same
predictions. Therefore, the prediction accuracy for LP index is not
sensitive to the parameter $\epsilon$ when $\epsilon$ is not so
large. Indeed, setting $\epsilon$ as a small positive number like
$0.01$ one can obtain a near optimal accuracy, usually less than
$1\%$ smaller than the real optimum (see also Table II, where we
compare the optimal AUC values with the values obtained by setting
$\epsilon=0.01$ for the six real networks). In finding the optimal
value of $\beta$, one can first calculate the maximal eigenvalue of
the adjacency matrix, and the optimal $\beta$ is always smaller than
its reciprocal. It is then easy to approach the optimal $\beta$. For
example, the optimal values of $\beta$ for relevant data points
shown in Fig. 1 and Fig. 2 are all no larger than 0.03.

Next, we discuss the computational complexities of the three
similarity indices. In calculating the CN index, for each node,
denoted by $x$, we first search all $x$'s neighbors (called the step
1), and then lay out the neighbors of each of $x$'s neighbors,
respectively (called the step 2). If a node $y$ appears $n$ times in
the step 2, $s_{xy}=n$. Since the time complexity to traverse the
neighborhood of a node is simply $k$, the time complexity in
calculating CN index is $\mathbb{O}(Nk^2)$. Analogously, for LP
index, what we need to do is go one step further (called the step 3)
to check all neighbors of each of $x$'s second-order neighbors,
respectively. If a node $y$ appears $n$ times in $x$'s second-order
neighborhood and $m$ times in $x$'s third-order neighborhood,
$s_{xy}=n+\epsilon m$. Therefore, the time complexity in calculating
the LP index is $\mathbb{O}(Nk^3)$. An detailed illustration for an
example network consisted of four nodes is shown in Fig.~\ref{Time}.
For the Katz index, the time complexity is mainly determined by the
matrix inversion operator, which is $\mathbb{O}(N^3)$
\cite{Golub1996}. In Fig. 4 and Fig. 5, we report the numerical
results about computational complexity of the three similarity
indices, which are well in accordance with the analysis. Beside the
time complexity, memory space is another limitation for algorithmic
implementation for huge-size networks. In calculating CN and LP
indices, the memory required are of the order $\mathbb{O}(Nk)$,
while for the Katz index, it is of the order $\mathbb{O}(N^2)$. In a
word, compared with the widely applied CN index and Katz index, the
LP index is not only highly effective (i.e., accurate), but also
highly efficient (i.e., required relatively less memory and CPU
time).

\section{Empirical Analysis}

In this paper, we consider six representative networks drawn from
disparate fields: (i) PPI.--- A protein-protein interaction network
containing 2617 proteins and 11855 interactions \cite{Mering2002}.
Although this network is not connected (it contains 92 components),
most of nodes belong to the giant component, whose size is 2375.
(ii) NS.--- A network of coauthorships between scientists who are
themselves publishing on the topic of networks \cite{Newman2006}.
This network contains 1589 scientists, and 128 of which are
isolated. Here we do not consider those isolated nodes. The
connectivity of NS is not good. It is consisted of 268 connected
components, and the size of the largest connected component is only
379. (iii) Grid.--- An electrical power grid of western US
\cite{Watts1998}, with nodes representing generators, transformers
and substations, and links corresponding to the high voltage
transmission lines between them. This network contains 4941 nodes
and is well connected. (iv) PB.--- A network of the US political
blogs \cite{Ackland2005}. The original links are directed, here we
treat them as undirected ones. PB has 1224 nodes and the giant
component contains 1222 nodes. (v) INT.--- The router-level topology
of the Internet, which is collected by the \emph{Rocketfuel Project}
\cite{Spring2004}. INT has 5022 nodes and is well connected, while
it is an extremely sparse network with average degree being only
2.49. (vi) USAir.--- the network of US air transportation system,
which contains 332 airports and 2126 airlines \cite{Pajek}. Note
that, all the similarity indices considered here, as well as those
well-known indices (except the preferential attachment index)
reported in Refs. \cite{Kleinberg2007,Linyuan2009}, will give zero
score to a pair of nodes located in two disconnected components.
Therefore, here we only consider the giant component, and when
preparing the probe set, we also make sure that the remain training
set representing a connected network. Actually, each time before
removing of a link to the probe set, we first check if this removal
will make the training network disconnected. Table 1 summarizes the
basic topological features of the giant component of those networks.
Brief definitions of the monitored topological measures can be found
in the table caption, for more details, please see the review
articles
\cite{Albert2002,Dorogovtsev2002,Newman2003,Boccaletti2006,Costa2007}.

\begin{table}
\caption{The basic topological features of the giant components of
the six example networks. $N$ and $M$ are the total numbers of nodes
and links, respectively. $\langle{k}\rangle$ is the average degree
of the network. $\langle{d\rangle}$ is the average shortest distance
between node pairs. $C$ and $r$ are clustering coefficient
\cite{Watts1998} and assortative coefficient \cite{Newman2002},
respectively. Nodes with degree 1 are excluded from the calculation
of clustering coefficient. $H$ is the degree heterogeneity, defined
as $H=\frac{\langle k^2\rangle}{\langle k\rangle^2}$, where $\langle
k\rangle$ denotes the average degree.}
\begin{center}
\begin{tabular} {cccccccc}
  \hline \hline
   Networks     & $N$  &  $M$  &  $\langle k\rangle$ & $\langle d\rangle$ & $C$ & $r$ & $H$ \\
   \hline
   PPI & 2375 & 11693 & 9.847 & 4.59 & 0.388 & 0.454 & 3.476 \\
   NS & 379 & 941 & 4.823 & 4.93 & 0.798 & -0.082 & 1.663 \\
   Grid & 4941 & 6594 & 2.669 & 15.87 & 0.107  & 0.003 & 1.450 \\
   PB & 1222 & 16717 & 27.360 & 2.51 & 0.360  & -0.221 & 2.970 \\
   INT & 5022 & 6258 & 2.492 & 5.99 & 0.033  & -0.138 & 5.503 \\
   USAir & 332 & 2126 & 12.807 & 2.46 & 0.749  & -0.208 & 3.464 \\
   \hline \hline
    \end{tabular}
\end{center}
\end{table}

We apply the link prediction algorithm on the six real networks, and
the accuracies is shown in Table 2, with those entries corresponding
to the highest accuracies being emphasized by black. Clearly, the LP
index always performs better than the CN index, especially, for INT,
the AUC is sharply improved from 0.653 to 0.943. Except Grid, the LP
index gives competitively accurate predictions as the Katz index.
Grid is a strongly localized network with most of links being of
short geographical lengths, and thus the average topological
distance of Grid, $\langle d\rangle=15.87$, is much larger than the
other five example networks. Although Grid is geographically
localized, the clustering coefficient is relatively small and it
lacks short loops since such loops are redundant and of lower
efficiency in the engineering viewpoint. Actually, in Grid, when a
link is removed, it is usually hard to find a very short path (like
of length 2 or 3) connecting the two endpoints. Therefore, the CN
and LP indices, considering only very short paths, fail to re-find
the correlation between two directly connected nodes if the link is
removed. In addition, we note that the optimal value of $\epsilon$
for USAir is negative. In USAir, the large-degree nodes are densely
connected and share many common neighbors. Even without the
contribution of $\epsilon A^3$, the links among large-degree nodes
are assigned very high scores, thus the additional item, $\epsilon
A^3$, changes little of their relative positions. Considering two
small local airports, $x$ and $y$, which are connected to their
local central airports, $x'$ and $y'$. Of course, many hubs are
common neighbors of $x'$ and $y'$, and $x'$ and $y'$ may be directly
connected. If the link $(x,x')$ is removed, the similarities between
$x$ and other nodes are all zero. Otherwise, the similarities
$s_{xy'}$ (by $x$-$x'$-hub-$y'$), $s_{xy}$ (by $x$-$x'$-$y'$-$y$),
and $s_{xh}$ where $h$ represents a hub node (by $x$-$x'$-hub-$h$ or
$x$-$x'$-$y'$-$h$) are positive due to the contributions of paths
with length 3. There are many links connecting small local airports
and local centers, some of which are removed, and the others are
kept in the testing set. According to the above discussion, the
removed links have lower score than the nonexistent links due to the
additional item $\epsilon A^3$. In a word, the very specific
structure of USAir (the hierarchical organization consisted of hubs,
local centers and small local airports) makes the LP index with
positive $\epsilon$ worse than the simple CN corresponding to
$\epsilon=0$, which is also the reason why negative $\epsilon$
performs even better.

\begin{table}
\caption{Accuracies of the three similarity indices, measured by the
area under the ROC curve (AUC). Each number is obtained by averaging
over 10 independent realizations. The entries corresponding to the
highest accuracies are emphasized by black. For LP and Katz indices,
the AUC values are corresponding to the optimal parameter. LP*
denotes the LP index with a fixed parameter $\epsilon=0.01$. The
very small difference between the optimal case and the case with
$\epsilon=0.01$ suggests that in the real application, one can
directly set $\epsilon$ as a very small number, instead of finding
out its optimum that may cost much time.}
\begin{center}
\begin{tabular} {cccccccc}
  \hline \hline
   Nets     & PPI  &  NS  &  Grid & PB & INT & USAir \\
   \hline
   CN & 0.915 & 0.983 & 0.627 & 0.924 & 0.653 & 0.958  \\
   LP & 0.970 & \textbf{0.988} & 0.697 & \textbf{0.941} & 0.943 &
   \textbf{0.960}\footnote{For USAir, the optimal value of $\epsilon$ is negative. See the explanation in text.} \\
   Katz & \textbf{0.972} & \textbf{0.988} & \textbf{0.952} & 0.936 & \textbf{0.975} & 0.956  \\
   LP* & 0.970 & 0.988 & 0.697 & 0.939 & 0.941 & 0.959\footnote{For USAir, we set $\epsilon=-0.01$.}\\
   \hline \hline
    \end{tabular}
\end{center}
\end{table}

Table 3 presents the computation time of the link prediction
algorithm on the three similarity indices. Clearly, CN costs the
least. Note that, the computational complexity in calculating the LP
index is very sensitive to the average degree, while the one in
calculating the Katz index is very sensitive to the network size.
Therefore, the algorithm using LP index has great superiority for
the huge-size and sparse networks compared with the one adopting the
Katz index. Take INT as an example, the algorithm using the Katz
index runs about one day while the one using the LP index takes less
than half minute. Since the real challenge on computational
complexity is always relevant to the huge-size real networks, which
are mostly very sparse \cite{Albert2002}, the LP index is much more
practical than the Katz index. As a final remark, one may concern
that whether to employ higher-order paths is worthwhile in practice,
like to define a similarity index in the form
\begin{equation}
S=A^2+\epsilon A^3+\epsilon^2 A^4.
\end{equation}
We give a brief discussion on this issue in \emph{Appendix A}.

\begin{table}
\caption{Computation time (in microsecond) of the link prediction
algorithm on the three similarity indices of the six example
networks. The hardware environment is the same as what we stated in
the caption of Figure 4.}
\begin{center}
\begin{tabular} {cccccccc}
  \hline \hline
   Nets     & PPI  &  NS  &  Grid & PB & INT & USAir \\
   \hline
   CN & 10690 & 253 & 5161 & 31112 & 6711 & 2208 \\
   LP & 543589 & 1638 & 11344 & 2873403 & 27641 & 93892  \\
   Katz & 8073316 & 27479 & 69961063 & 1051528 & 72550935 & 17603  \\
   \hline \hline
    \end{tabular}
\end{center}
\end{table}

\begin{table}
\caption{Comparison of the accuracies of the original local path
index ($n=3$, see Eq. (3)) and the higher-order local path index
($n=4$, see Eq. (5)), measured by the area under the ROC curve
(AUC). Each number is obtained by averaging over 10 independent
realizations. The AUC values reported here are corresponding to the
optimal parameter. The average shortest distance and the improvement
(\%) by considering higher-order paths are also laid out in this
Table, and all the six real networks are ordered by their shortest
average distances.}
\begin{center}
\begin{tabular} {cccccccc}
  \hline \hline
   Nets     & USAir  &  PB  &  PPI & NS & INT & Grid \\
   \hline
   $\langle d \rangle$ & 2.46 & 2.51 & 4.59 & 4.93 & 5.99 & 15.87  \\
   $n=3$ & 0.960 & 0.941 & 0.970 & 0.988 & 0.943 & 0.697 \\
   $n=4$ & 0.959 & 0.937 & 0.973 & 0.989 & 0.959 & 0.759  \\
   Improvement  & -0.104 & -0.425 & 0.309 & 0.101 & 1.70 & 8.90 \\
   \hline \hline
    \end{tabular}
\end{center}
\end{table}

\section{Conclusion and discussion}

In this paper, we introduced a local path index to estimate the
likelihood of the existence of a link between two nodes. We propose
a network model with controllable density and noise strength in
generating links. The LP index provides slightly more accurate
predictions than the Katz index, especially in the highly noisy
cases. We further use six representative real networks to test the
three similarity indices, showing that the LP index can provide
competitively accurate predictions as the Katz index. Compared with
the Katz index, the LP index requires much less CPU time and memory
space, and is therefore more practical. Ignored the degree-degree
correlation, the time complexities in calculating LP index and Katz
index are $\mathbb{O}(N\langle k\rangle^3)$ and $\mathbb{O}(N^3)$,
respectively. Hence for the huge (i.e., very large $N$) and sparse
(i.e., very small average degree $\langle k\rangle$) networks, the
advantage of the LP index is striking.

Highly accurate predictions are significant in practice. For
example, many biological networks, such as protein-protein
interaction networks, metabolic networks and food webs, the
discovery of links/interactions costs much in the laboratory or the
field. Instead of blindly checking all possible interactions, to
predict in advance based on the interactions known already and focus
on those links most likely to exist can sharply reduce the
experimental costs if the predictions are accurate enough
\cite{Newman2008Nature,Redner2008}. For some others like the
friendship networks in web society, very likely but not yet existent
links can be suggested to the relevant users as recommendations of
promising friendships. These recommendations can help users finding
new friends and thus enhance their loyalties to the web sites.
Besides the practical significance, it is worthwhile to emphasize
that the study of link prediction can also provide some theoretical
insights about the structural organization. For example, in this
paper, the unexpected results on Grid and USAir give evidence to
some specific structural properties that are not straightforwardly
notable. Another example is that the preferential attachment index
usually gives poor predictions, and when it works relatively good,
it implies that the testing network has strong rich-club phenomenon
\cite{Linyuan2009,Huang2005}. Although the focus of this paper is
not to investigate the relations between suitable similarity indices
and network structures, we believe it is an interesting issue worth
further studies.

In this paper, we only considered the link prediction problem in
static networks. However, many real networks are evolving all the
time, and the links created in different times should be assigned
different weights in principle. This time-involved link prediction
problem is rarely investigated and of course worths a serious study
in the future \cite{Liu2009}. Most of previous studies in relevant
direction only test the algorithmic accuracy in real networks. Here
we argue that the modeled networks should be used, because one can
control some meaningful parameters in a model, which can not be
directly observed in the real networks (e.g., the strength of noise
or irrationality). We hope the proposed model could become a
prototype in testing the accuracy of link prediction algorithms,
however, it is currently too simple and to make it closer to the
real networks, such as introducing controllable degree heterogeneity
and degree-degree correlation, is very helpful.

This paper concerns only the simple networks, however, the local
path index can be easily extended to more complicated cases. For
example, we can handle the directed networks by replacing the
original adjacency matrix, $A$, by an asymmetry one, the weighted
networks by replacing $A$ by a weighted matrix, and the networks
with self connections by assigning nonzero diagonal elements.
Actually, Murate and Moriyasu \cite{Murata2007} have already
investigated the link prediction problem in weighted networks,
however, the credibility of their work is recently challenged by the
empirical evidence that the weak ties may play a more important role
in link prediction than the strong ties \cite{Linyuan2009b}.

\begin{acknowledgments}
The authors acknowledge valuable discussion with Yi-Cheng Zhang.
This work is partially supported by the Swiss National Science
Foundation (Project 205120-113842). C.-H.J. acknowledges the Future
and Emerging Technologies programmes of the European Commission
FP7-COSI-ICT (Project QLectives, Grant No. 231200). T.Z.
acknowledges the National Natural Science Foundation of China (Grant
Nos. 10635040 and 60744003).
\end{acknowledgments}

\begin{appendix}

\section{Similarity Index Involving Higher-Order Paths}

A straightforward method to extend the local path index is to
consider the higher-order paths. Such a similarity index is of the
form
\begin{equation}
S=A^2+\epsilon A^3+\epsilon^2 A^4+\cdots+\epsilon^{n-2} A^n,
\end{equation}
where $n>2$ is the maximal order. As shown in Fig. 3, the
computational complexity in an uncorrelated network is
$\mathbb{O}(N\langle k\rangle^n)$, which grows fast with the
increasing of $n$ and will exceed the complexity for calculating the
Katz index for large $n$. We therefore concentrate on the case of
$n=4$, equivalent to the one shown in Eq. (5).

As shown in Table IV, the improvements of accuracy are not much
except for the power grid. Sometimes, to introduce higher-order
relations will even decrease the accuracy, like for USAir and PB.
The results are very sensitive to the average shortest distances of
networks. If $\langle d\rangle$ is very short, to consider paths
with length three seems enough, and the addition item, $\epsilon^2
A^4$, will make little effort (e.g., PPI, NS and INT) or even
negative effort (e.g., USAir and PB). Only when the network is of
long average shortest distance, to consider higher-order relations
may be cost-effective. Since most real networks exhibit strongly
small-world effect
\cite{Albert2002,Dorogovtsev2002,Newman2003,Boccaletti2006,Costa2007},
a local path index taking into account paths with length no more
than three may be practically sufficient.

\end{appendix}

\end{document}